# Microwave Power-to-Frequency Transduction via Injection Pulling of a Self-Sustained Oscillator for Rydberg Superheterodyne Sensing

D. Arumugam

*Jet Propulsion Laboratory, California Institute of Technology*

(Electronic mail: darmindra.d.arumugam@jpl.nasa.gov.)

A Rydberg superheterodyne sensing architecture is demonstrated in which a self-sustained oscillator (SSO) serves as a dynamically perturbed local oscillator (LO) for microwave detection. The SSO is realized by a phase-controlled radio-frequency (RF) feedback loop coupled to a transverse electromagnetic (TEM) cavity containing a Rydberg vapor cell. The system operates near 5.49 GHz using a cesium ladder scheme with an 852 nm probe and 510 nm coupling laser addressing the $6\mathrm{S}_{1/2}$–$6\mathrm{P}_{3/2}$–$49\mathrm{D}_{3/2}$ transition, with microwave coupling to the $50\mathrm{P}_{3/2}$ state. Injection of a microwave signal pulls the SSO frequency via nonlinear dynamics, converting input power into a measurable frequency shift read out optically as a Rydberg probe intermediate-frequency (IF) signal. The response follows Adler-type injection-pulling behavior, with continuous IF tuning with input power. A peak responsivity of ~35 kHz/dB is observed, with enhanced sensitivity near synchronization. These results demonstrate power-to-frequency transduction using a dynamically perturbed LO combined with Rydberg atomic readout.

Microwave and radio-frequency (RF) sensing underpins radar, communications, and remote sensing [1–3]. Local-oscillator (LO)-based superheterodyne architectures [3,4] are widely used for RF detection, enabling sensitive signal extraction. Detection of weak RF signals remains limited by noise and LO stability in conventional receiver architectures [2-3,5], motivating the widespread use of externally supplied, highly stable oscillators. Rydberg atoms enable highly sensitive RF electric-field (E-field) detection through their large dipole moments and tunable resonances [6–10]. These systems have demonstrated broadband and calibration-free sensing capabilities [11,12], and are typically operated in linear regimes where the measured response scales with the applied field amplitude [11]. This includes implementations based on electromagnetically induced transparency (EIT) and Autler–Townes splitting (ATS) [9]. Rydberg superheterodyne detection has also been demonstrated using an applied LO field for sensitive E-field detection [13,14], where the LO acts as an externally supplied, fixed reference, similar to classical architectures. Separately, injection pulling and synchronization have been extensively studied in nonlinear oscillators, but primarily in purely classical systems without atomic readout [15,16]. Recent work in nonlinear atomic regimes, including Rydberg dissipative time crystals [17], has demonstrated enhanced sensitivity through dynamical responses of the atomic medium [18]. In these systems, injection-pulling dynamics near synchronization convert small perturbations into measurable frequency shifts and exhibit strong sensitivity to external signals [19]. These results suggest that nonlinearity can provide new sensing modalities when appropriately exploited.

Nonlinear oscillators coupled to Rydberg atom sensors could provide an alternative sensing mechanism; however, this approach has not yet been explored. In particular, a self-sustained LO in a Rydberg superheterodyne system has not been demonstrated. In such a system, the LO frequency would be dynamically shifted toward an incident microwave signal via injection-pulling dynamics [15,16], converting incident microwave power into a measurable LO frequency shift. This enables transduction through the nonlinear response of the oscillator, while preserving atomic susceptibility as a linear and high-sensitivity readout mechanism. Such an architecture would fundamentally differ from existing approaches, in which the LO serves as a fixed reference rather than actively responding to microwaves.

Here, a hybrid classical–atomic sensing architecture is demonstrated in which a self-sustained oscillator (SSO) acts as a dynamically perturbed LO coupled to a Rydberg medium. Injection of a microwave signal to be sensed, pulls the SSO frequency, which is transduced into an optical intermediate-frequency (IF) signal via the Rydberg response, enabling power-to-frequency conversion of the microwave signal.

## I. HYBRID CLASSICAL-ATOMIC SYSTEM AND EXPERIMENT

The experimental platform is designed to realize a microwave sensing modality based on power-to-frequency transduction, in which an incident RF field is encoded as a frequency shift of a LO for superheterodyne-style readout using a Rydberg atomic medium. To achieve this, a hybrid classical–atomic system is implemented in which a microwave SSO is coupled to a broadband transverse electromagnetic (TEM) cavity (commercial Open TEM TBTC0, DC–6 GHz) containing a room-temperature cesium (Cs) vapor cell (Fig. 1a). In contrast to conventional superheterodyne receivers that rely on an externally supplied LO, the present approach generates internal self-sustained RF oscillations whose frequency is directly perturbed by the incident microwave field via injection pulling, enabling direct transduction of input power into frequency excursions. The RF loop comprises a low-noise amplifier (Mini-Circuits ZX60-5916M-S, 1.5–5.9 GHz), a circulator (DITOM D3C2080, 2–8

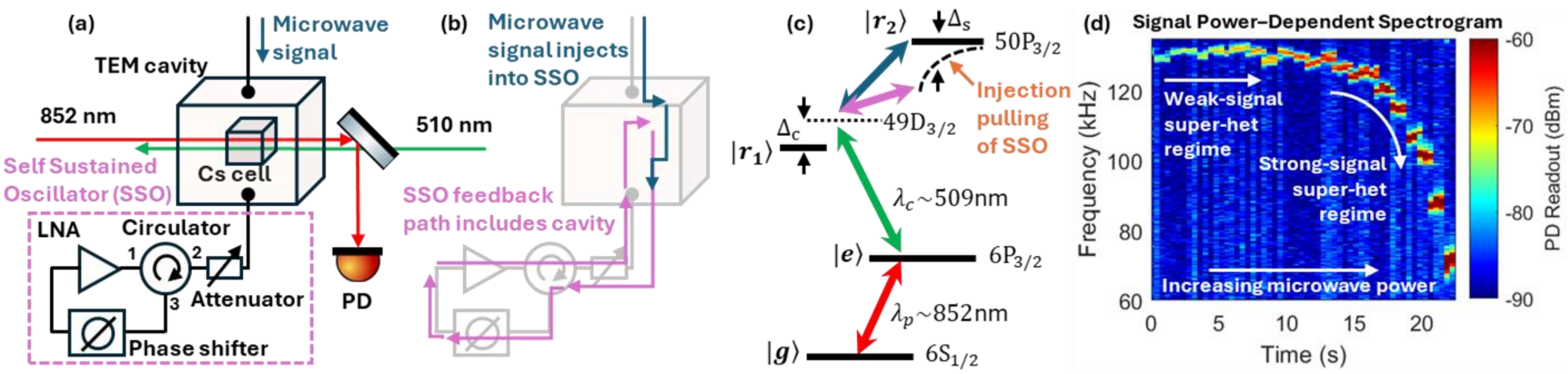

**Figure 1. Rydberg-coupled SSO and injection-pulling response.** (a) System schematic: A Cs vapor cell inside a microwave TEM cavity is embedded in an RF feedback loop (LNA + circulator + phase control), forming a self-sustained oscillator whose frequency is set by the cavity response and phase shifters. Optical probe transmission via a 2-photon Rydberg system provides readout. (b) Feedback architecture: The cavity response, including internal reflections, is part of the loop phase condition. A microwave signal is injected through a separate port, perturbing the free-running oscillator and enabling injection pulling. (c) Rydberg excitation scheme: A ladder system in Cs is used, with a probe laser near 852 nm driving the $6S_{1/2} \rightarrow 6P_{3/2}$ transition and a coupling laser near 510 nm driving $6P_{3/2} \rightarrow 49D_{3/2}$ state. The microwave field couples nearby $50P_{3/2}$ Rydberg state, and the SSO field is detuned by $\Delta_s$ – which is dynamically modified by the microwave field. (d) Signal power–dependent spectrogram of the photodetector readout. At low microwave injected power, the system operates in a weak-perturbation (superheterodyne) regime with minimal frequency shift. With increasing power, the oscillator enters a nonlinear injection-pulling regime with large frequency excursions, approaching a strong-coupling regime at the highest powers.

GHz) to enforce unidirectional propagation, two serially connected continuously tunable phase shifters (Weinschel 981, DC–18 GHz; Weinschel 980-4K, DC–12 GHz), and a programmable attenuator (Mini-Circuits RCDAT-8000-30, 1–8 GHz) for gain control. Self-sustained oscillation arises when the loop satisfies the Barkhausen condition, with the oscillation frequency determined primarily by the total loop phase, which can be continuously tuned via the phase shifters. In this configuration, the open TEM cavity acts as a broadband interaction region rather than a high-Q frequency-selective element, and the Cs vapor cell—being small relative to the cavity volume—introduces only a weak perturbation to the RF boundary conditions. As a result, the free-running oscillator frequency is governed predominantly by the electronic loop phase response, while the cavity provides a controlled region for coupling the RF field to the atomic medium. A microwave signal to be sensed, is injected into the cavity through an independent port, perturbing the steady-state oscillation and enabling controlled access to injection-pulling dynamics, which form the basis of the power-to-frequency transduction mechanism (see Fig. 1b).

The atomic readout is implemented using a ladder-type Rydberg excitation scheme in Cs within the vapor cell located in the TEM cavity interaction region (Fig. 1c). The optical transitions follow the $6S_{1/2} \rightarrow 6P_{3/2} \rightarrow 49D_{3/2}$ pathway, with a probe laser near 852 nm addressing the $6S_{1/2},\ F = 4 \rightarrow 6P_{3/2},\ F' = 4/5$ crossover transition and a coupling laser near 510 nm driving the $6P_{3/2} \rightarrow 49D_{3/2}$ Rydberg transition. Both probe and coupling beams are linearly polarized and co-propagating through the vapor cell, with beam diameters of approximately 1.2mm (1/e$^2$). The probe laser is frequency-stabilized via Doppler-free saturation absorption spectroscopy in a reference Cs cell, yielding a linewidth of <78kHz. The coupling laser is stabilized using a Pound–Drever–Hall (PDH) lock to a high-finesse ultra-low expansion (ULE) reference cavity (finesse $\sim 1.5 \times 10^4$), providing sub-kHz long duration frequency stability required for consistent Rydberg excitation. The 510 nm light is generated via second-harmonic conversion of a near-infrared fundamental at 1020 nm using a nonlinear crystal-based frequency-doubling stage.

Microwave coupling is established between the $49D_{3/2}$ state and a nearby $50P_{3/2}$ Rydberg state, allowing the intracavity RF field—set by the self-sustained oscillator and modified by injected signals—to directly perturb the atomic susceptibility. Since the SSO, which acts as an LO is pulled, the weak microwave field to be sensed in tuned to on resonance $49D_{3/2} \rightarrow 50P_{3/2}$. Optical detection is performed using a balanced detection scheme in which the probe beam is split into signal and reference arms. The signal beam co-propagates with the coupling beam through the vapor cell and experiences the full Rydberg interaction, while the reference beam traverses the cell without overlap with the coupling field, thereby sampling only the baseline absorption and common-mode technical noise. The differential photodetector output suppresses intensity noise and enhances sensitivity to Rydberg-induced transmission changes associated with the microwave field. The resulting signal is recorded and analyzed in both time and frequency domains, with RF electronic passband over a bandwidth of 10–250 kHz (3dB passband) to isolate the relevant Rydberg superheterodyne and injection-pulling dynamics.

The power-dependent spectrogram of the Rydberg IF readout (Fig. 1d) shows the evolution of the oscillator with temporally increasing injected microwave power. At low power, the system operates in a linear superheterodyne regime, where the IF is set by the detuning between the free-running oscillator and injected signal ($\Delta_s$), and the readout amplitude increases with the injected microwave power. With increasing power, nonlinear injection pulling shifts the oscillation frequency toward the injected tone, reducing the IF. At higher powers, the system approaches synchronization, with the IF collapsing toward zero. This transition reflects the conversion from perturbative mixing to nonlinear power-to-frequency transduction observed in the optical readout.

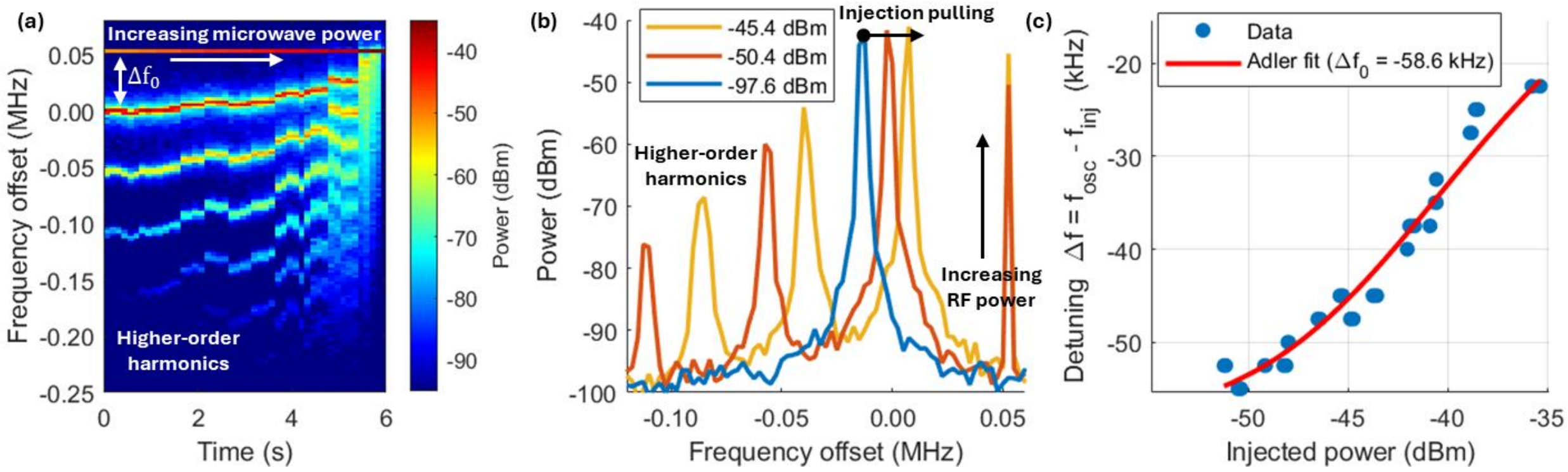

**Figure 2. Characterization of the SSO under microwave injection.** (a) Time-resolved spectrogram of the SSO output measured via a directional coupler tap, showing frequency offset relative to a driven microwave field placed at 5.489 GHz (the on-resonant Rydberg transition for $49D_{3/2} \rightarrow 50P_{3/2}$) as injected microwave power increases. The free-running oscillator exhibits a stable frequency at low injection $(\Delta f_0)$, with higher-order harmonic components visible. Increasing injected power produces a continuous frequency shift due to injection pulling. (b) Representative spectra at different injected powers, showing the evolution from a narrow free-running peak to progressively shifted peaks under increasing microwave injection. Higher-order harmonics are present under injection. The primary oscillation peak shifts toward the injected frequency as power increases. (c) Extracted detuning $\Delta f = f_{osc} - f_{inj}$ as a function of injected microwave power. The data follow an Adler-type dependence, confirming injection-pulling behavior of the SSO under external microwave perturbation.

Time-resolved characterization of the SSO under microwave injection is performed by tapping the intracavity field with a broadband directional coupler (Mini-Circuits ZCDZ30-5R263-S+, 0.5–26.5 GHz) placed between the SSO loop and the TEM cavity, with the coupled signal analyzed using an RF spectrum analyzer. The spectrogram in Fig. 2a shows the oscillator frequency referenced to the injected tone as its power is increased in time. The free-running offset $\Delta_s = \Delta f_0$ denotes detuning between the intrinsic SSO frequency and the injected signal in the absence of coupling. At low injected power, the oscillator remains near $\Delta f_0$ with minimal drift, while higher-order harmonic components appear in the presence of the injected signal, indicating mixing between the free-running oscillation and the external drive (Fig. 2a,b).

Representative spectra at increasing injection levels, shown in Fig. 2b, exhibit a narrow free-running peak ($f_{\mathrm{osc}}$) that shifts continuously toward the injected frequency ($f_{\mathrm{inj}}$) with increasing power, demonstrating injection pulling while maintaining nearly constant amplitude. Concurrently, additional harmonic components emerge only in the presence of the injected signal, consistent with nonlinear mixing between the oscillator and external field. The extracted detuning $\Delta f = f_{\mathrm{osc}} - f_{\mathrm{inj}}$ in Fig. 2c decreases monotonically with injected power, approaching zero without amplitude suppression, indicating a phase-dominated response of the limit-cycle oscillator. This behavior is consistent with Adler-type phase dynamics, $\dot{\phi} = \Delta\omega_0 - \kappa \sin\,\phi$, where the effective coupling scales with the injected field amplitude, $\kappa \propto \sqrt{P_{\mathrm{inj}}}$. In contrast to the ideal Adler solution, which predicts a sharp locking threshold, the observed response exhibits a smooth approach to synchronization, reflecting non-idealities such as finite linewidth, loop filtering, and amplitude–phase coupling. The data in Fig. 2c is well described by the phenomenological form $\Delta f(P_{\mathrm{inj}}) = (\Delta f_0)/\sqrt{(1 + \beta P_{\mathrm{inj}})}$, which captures the experimentally observed continuous reduction of detuning with injected power and provides an effective description of the softened injection-pulling dynamics. Here, $\beta$ represents a fitting parameter for this phenomenological Adler form.

## II. RYDBERG ATOMIC READOUT OF INJECTION PULLING DYNAMICS

Spectrograms of the Rydberg probe readout, shown in Fig. 3a,b, directly capture the IF response generated by the atoms under an injection-pulled LO. In this configuration, the injected microwave to be sensed drives the $49D_{3/2} \rightarrow 50P_{3/2}$ transition and the SSO serves as the LO close to the transition, while the optical probe transmission provides a down-converted measurement of the RF field through atomic susceptibility. The frequency axis therefore represents the Rydberg-derived IF signal rather than the raw RF spectrum, enabling direct optical readout of the oscillator dynamics. As the injected microwave power increases, the LO frequency is progressively pulled toward the external signal, and this evolution is faithfully transduced by the atoms as a continuous shift of the IF trajectory. The two cases correspond to different initial free-running detunings, $\Delta f_0 \approx 118$ kHz and $\Delta f_0 \approx 131$ kHz, which set the starting offset of the Rydberg readout. In both cases, the spectrograms exhibit a smooth, power-dependent reduction of the IF toward lower frequencies, consistent with injection-pulling behavior observed in the RF domain, but here measured entirely through the vapor-cell-based optical detection. This demonstrates that the Rydberg medium functions as a coherent superheterodyne transducer, converting microwave-induced power-to-frequency shifts of the LO into a directly observable probe optical signal. Notably, this behavior is not restricted to a narrowly tuned initial detuning, but is reproducible across a range of $\Delta f_0$, indicating that the transduction mechanism is robust provided the oscillator detuning remains within the effective detection bandwidth of the Rydberg sensor. In the present atomic system, this bandwidth is set primarily by the Rydberg resonance linewidth, determined by the coupling-laser Rabi frequency and associated power and decoherence broadening, allowing the atoms to track the injection-pulled oscillator over a finite detuning range.

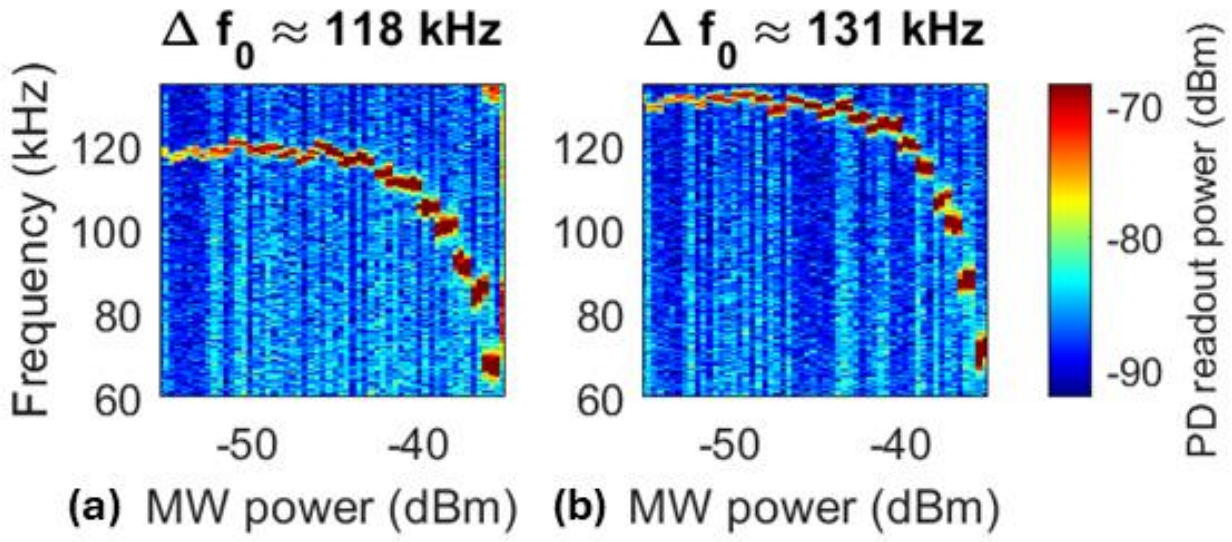

**Figure 3. Rydberg probe spectrogram under an injection-pulled LO.** (a–b) Spectrograms of the optical probe readout from a Cs Rydberg ladder system as a function of injected microwave power. The LO is a SSO that drives the atoms near the $49D_{3/2} \rightarrow 50P_{3/2}$ transition (5.489 GHz) and is progressively pulled by an externally injected microwave signal. The frequency axis is the baseband down-converted readout by the atoms. The two cases correspond to different free-running SSO LO detunings, $\Delta f_0 \approx 118$ kHz (a) and $\Delta f_0 \approx 131$ kHz (b). As the injected power increases, the LO frequency—and thus the Rydberg response—shifts continuously toward the injected signal, producing a characteristic injection-pulling trajectory in the probe readout.

The peak frequency extracted from the Rydberg IF readout (from underlying data of Fig. 3b, $\Delta f_0 \approx 131$ kHz) is shown in Fig. 4 as a function of injected microwave power, providing a compact representation of the injection-pulling dynamics sensed directly by the atomic medium. The data are obtained by tracking the dominant Rydberg IF tone. Error bars denote the 3 dB linewidth of the extracted peak which accounts for noise including SSO frequency noise when injection pulled. This provides a measure of spectral broadening and uncertainty in peak identification. The resulting peak trajectory exhibits a monotonic shift toward lower frequencies with increasing injected power, consistent with injection pulling of the LO, but here observed entirely through the vapor-cell-based Rydberg optical detection. Fitting the data to the Adler-type phenomenological model provides a quantitative description of the power-dependent frequency shift, enabling direct comparison between the RF-domain oscillator dynamics (Fig. 2c) and their Rydberg-transduced optical readout (Fig. 4). The Adler-type model captures the overall pulling trend. However, unlike the direct RF measurement in Fig. 2c, the Rydberg IF readout is sensed through frequency-dependent atomic susceptibility and does exhibit some non-ideal deviations from the Adler form; as the SSO is pulled, changes in detuning from the target Rydberg transition can modify the optical response.

Building on the single-detuning characterization in Fig. 4, the injection-pulling dynamics are evaluated for two distinct initial detunings, $\Delta f_0 \approx 96$ kHz and $131$ kHz. The resulting trajectories in Fig. 5a show that both cases exhibit monotonic pulling toward the injected tone; however, the onset of strong frequency displacement occurs at lower injected power for the smaller detuning. This is consistent with Adler-type phase dynamics, $\dot{\phi} = \Delta\omega_0 - \kappa \sin\,\phi$, with $\kappa \propto \sqrt{P_{\text{inj}}}$, such that the condition $\kappa \sim \Delta\omega_0$ is reached at lower power for smaller $\Delta f_0$. Consequently, at the same injected power (e.g., near $-35$ dBm), the smaller-detuning case exhibits a significantly larger frequency shift.

The corresponding frequency responsivity $|\, df_{\text{out}}/dP_{\text{inj}}\,|$ in Fig. 5b quantifies this effect, reaching peak values of $\sim 35$ kHz/dB for $\Delta f_0 \approx 96$ kHz and $\sim 20$ kHz/dB for $\Delta f_0 \approx 131$ kHz at the highest injected powers. This demonstrates a substantial enhancement in transduction gain for smaller initial detuning, reflecting the increased phase sensitivity as the system approaches synchronization. Measurements at lower IF frequencies (<20 kHz) where even higher responsivity is expected near the critical boundary $\kappa \approx \Delta\omega_0$, are presently limited by residual PID control signals used to stabilize the probe laser to the hyperfine reference. These technical constraints restrict access to the immediate vicinity of the synchronization threshold, where the slope $df/dP$ is expected to increase further. Extending the measurement bandwidth into this regime represents an important direction for future work and is expected to yield significantly higher power-to-frequency transduction gain.

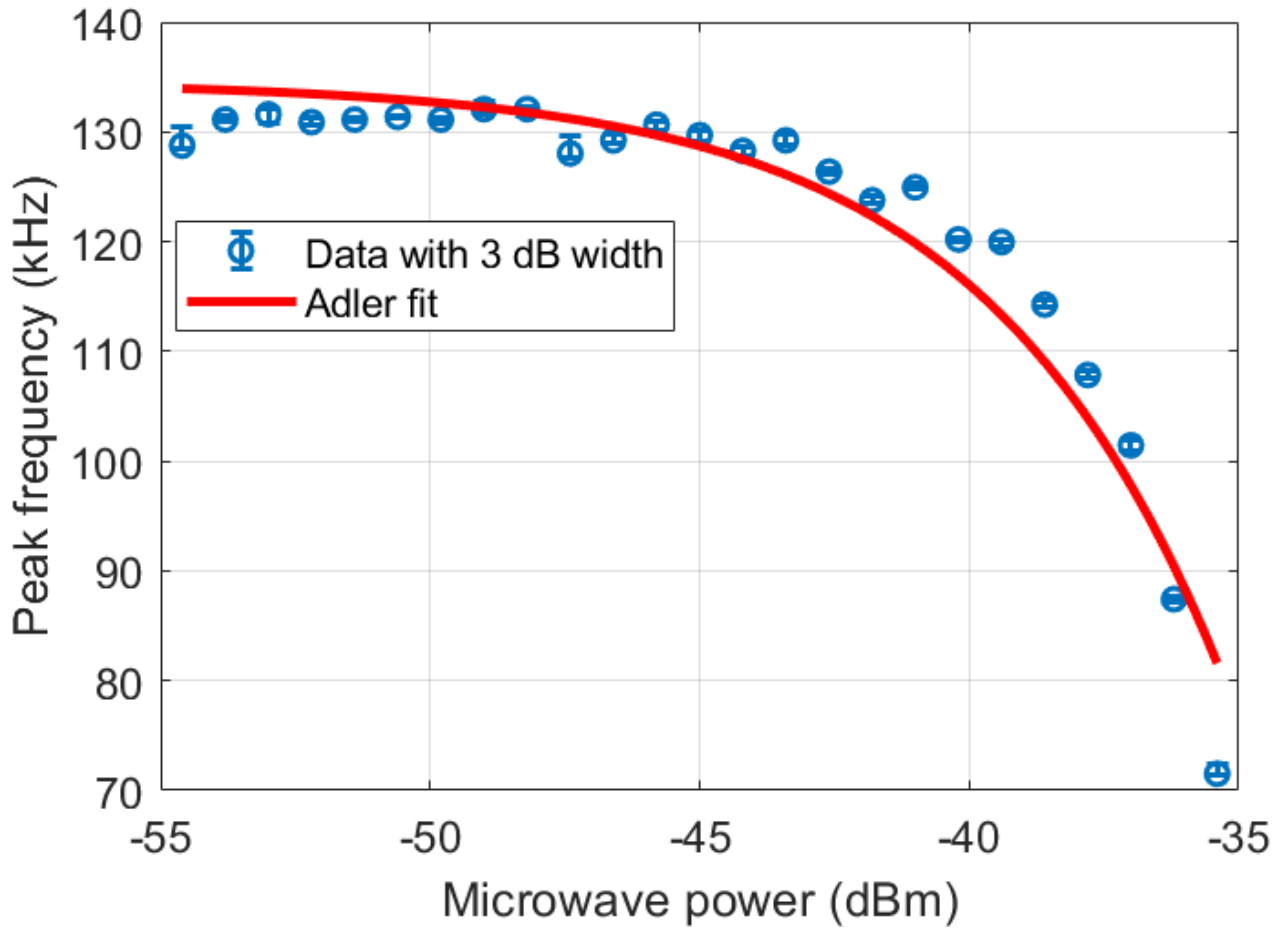

**Figure 4. Peak frequency shift of Rydberg baseband detection under microwave-induced injection pulling.** Peak frequency extracted from the Rydberg probe readout as a function of injected microwave power ($\Delta \mathrm{f}_0 \approx 131$ kHz). The LO, realized by a self-sustained oscillator driving the atoms near the $49D_{3/2} \rightarrow 50P_{3/2}$ transition (5.489 GHz), is progressively pulled by the injected signal, producing a monotonic frequency shift. Error bars denote the 3 dB linewidth of the extracted peak. The observed dependence follows an Adler-type model (red curve), consistent with injection-pulling dynamics, with the largest frequency shifts occurring at higher injected powers as the system approaches the strong-pulling regime.

These results demonstrate a hybrid classical–atomic sensing platform in which a microwave SSO coupled to a Rydberg sensing medium enables RF power to Rydberg IF frequency transduction via injection pulling. The oscillator exhibits Adler-type phase dynamics under external microwave drive, with a smooth approach to synchronization that is directly transduced into an optical IF signal through the atomic susceptibility. The Rydberg IF readout captures the nonlinear oscillator behavior while providing a fully optical superheterodyne measurement of RF dynamics. Quantitative analysis shows that the transduction gain is strongly dependent on the initial detuning, with enhanced responsivity achieved near the onset of synchronization. These results establish an injection-pulled SSO LO combined with Rydberg atomic superheterodyne readout as an effective mechanism for power-to-frequency transduction of RF signals.

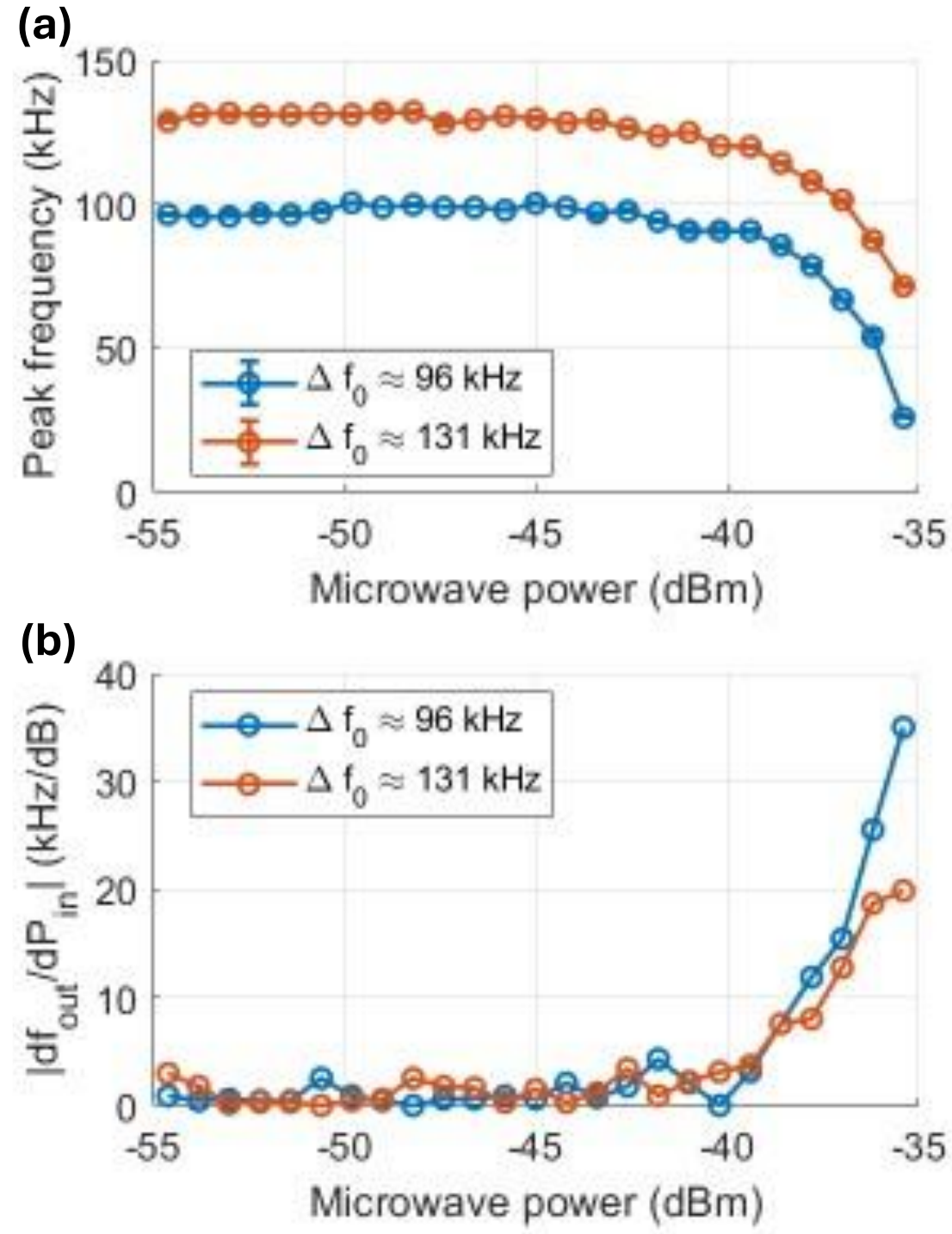

**Figure 5. Microwave power–to–frequency transduction via injection pulling for different initial detuning.** (a) Peak frequency extracted from the Rydberg probe (atomic IF) readout as a function of injected microwave power for two initial detunings, $\Delta f_0 \approx 96$ kHz and $\approx 131$ kHz. The LO, realized by a self-sustained oscillator driving the atoms near the $49D_{3/2} \rightarrow 50P_{3/2}$ transition (5.489 GHz), is progressively pulled toward the injected signal, producing a monotonic frequency shift. For smaller initial detuning, the onset of strong pulling occurs at lower injected power, resulting in a more rapid frequency shift at a given power level. (b) Corresponding frequency responsivity, $| df_{out}/dP_{in} |$, highlighting the nonlinear transduction gain. The responsivity increases sharply as the system approaches the strong-pulling regime, with significantly higher transduction gain achieved for smaller $\Delta f_0$ systems at the same power levels. This demonstrates that power-to-frequency conversion is strongly enhanced near the injection locking threshold, which is at lower powers for smaller $\Delta f_0$ systems.

ACKNOWLEDGMENTS

The research was carried out at the Jet Propulsion Laboratory, California Institute of Technology, under a contract with the National Aeronautics and Space Administration (80NM0018D0004), through the Instrument Incubator Program's (IIP) Instrument Concept Development (Task Order 80NM0022F0020).

AUTHOR CONTRIBUTIONS

D.A conceived the project and designed and implemented the experimental system, including the optical, microwave, and control subsystems, as well as the stabilization and locking architecture. D.A. developed the data acquisition and processing routines, performed the experiments, and analyzed the data. D.A. interpreted the results and prepared the manuscript.

DATA AVAILABILITY STATEMENT

The measurement spectrogram datasets corresponding to the measurements shown in Fig. 2a and Fig. 3b, including the processed power matrices and associated frequency and microwave power axes, are available as source data. All other data are available upon reasonable request to the corresponding author, Darmindra Arumugam via email: darmindra.d.arumugam@jpl.nasa.gov.